\documentclass[12pt, amsfonts]{article}
\begin{document}
\def\p {{\partial}}
\def\n {{\nu}}
\def\m {{\mu}}
\def\a {{\alpha}}
\def\bt {{\beta}}
\def\f {{\phi}}
\def\th {{\theta}}
\def\g {{\gamma}}
\def\eps {{\epsilon}}
\def\e {{\psi}}
\def\k {{\chi}}
\def\la {{\lambda}}
\def\na {{\nabla}}
\def\bn {\begin{eqnarray}}
\def\en {\end{eqnarray}}
\title{Canonical quantization of systems with time-dependent constraints\footnote{e-mail:
$sami_{-}muslih$@hotmail.com}} \maketitle
\begin{center}
\author{S.I. MUSLIH\\ {\it Deptartment of Physics, Al-Azhar University,
Gaza, Palestine}}
\end{center}

\begin{abstract}
The Hamilton - Jacobi method of constrained systems is discussed.
The equations of motion for a singular system with time dependent
constraints are obtained as total differential equations in many
variables. The integrability conditions for the relativistic
particle in a plane wave lead us to obtain the canonical phase
space coordinates with out using any gauge fixing condition. As a
result of the quantization, we get the Klein-Gordon theory for a
particle in a plane wave. The path integral quantization for this
system is obtained using the canonical path integral formulation
method.

\end{abstract}

\newpage

\section{Introduction}

There are many singular theories with constraints which depend on
time manifestly. The relativistic particle theories, string
theories and theories of gravity are some examples. The presence
of first class constraints in these theories forces one to impose
time dependent gauge fixing condition for each first class
constraint such that the whole set of constraints become
converted in to second-class constraints. The rules of canonical
quantization of systems with second-class constraints [1, 2],
which depend on time should be modified [3]. However, there exist
cases where one can make canonical transformation to make these
constraints time-independent and then use the ordinary rules of
quantization.

Recently, the canonical method [4-7] has been developed to
investigate constrained systems. The equations of motion are
obtained as total differential equations in many variables which
require the investigation of integrability conditions. If the
system is integrable, one can solve the equations of motion
without using any gauge fixing conditions.

In this paper we consider the quantization of the relativistic
spinless particle in the external field of a plane wave using the
Hamilton-Jacobi method [8-10]. in fact, this work is a
continuation of a previous work [11], where we have obtained the
path integral for a relativistic charged particle in an external
electromagnetic field and it is shown that the problems which
arise from identifying the measure of the path integral when
applying methods [1,2] and [12,13], are solved naturally by the
canonical path integral formulation [8-10].

 Now we would like to give a brief discussion of the canonical method.
This method gives the set of Hamilton - Jacobi partial
differential equations [HJPDE] as

\bn
&&H^{'}_{\a}(t_{\bt}, q_a, \frac{\p S}{\p q_a},\frac{\p S}{\p
t_a}) =0,\nonumber\\&&\a, \bt=0,n-r+1,...,n, a=1,...,n-r,\en where
\begin{equation}
H^{'}_{\a}=H_{\a}(t_{\bt}, q_a, p_a) + p_{\a},
\end{equation}
and $H_{0}$ is defined as
\bn
 &&H_{0}= p_{a}w_{a}+ p_{\m} \dot{q_{\m}}|_{p_{\n}=-H_{\n}}-
L(t, q_i, \dot{q_{\n}},
\dot{q_{a}}=w_a),\nonumber\\&&\m,~\n=n-r+1,...,n. \en

The equations of motion are obtained as total differential
equations in many variables as follows:

\bn
 &&dq_a=\frac{\p H^{'}_{\a}}{\p p_a}dt_{\a},\;
 dp_a= -\frac{\p H^{'}_{\a}}{\p q_a}dt_{\a},\;
dp_{\bt}= -\frac{\p H^{'}_{\a}}{\p t_{\bt}}dt_{\a}.\\
&& dz=(-H_{\a}+ p_a \frac{\p
H^{'}_{\a}}{\p p_a})dt_{\a};\\
&&\a, \bt=0,n-r+1,...,n, a=1,...,n-r\nonumber \en where
$z=S(t_{\a};q_a)$. The set of equations (4,5) is integrable [6,7]
if

\begin{equation}
dH^{'}_{0}=0,\;\;\; dH^{'}_{\m}=0,  \m=n-p+1,...,n.
\end{equation}
If condition (6) are not satisfied identically, one considers
them as new constraints and again testes the consistency
conditions. Hence, the canonical formulation leads to obtain the
set of canonical phase space coordinates $q_a$ and $p_a$ as
functions of $t_{\a}$, besides the canonical action integral is
obtained in terms of the canonical coordinates.The Hamiltonians
$H^{'}_{\a}$ are considered as the infinitesimal generators of
canonical transformations given by parameters $t_{\a}$
respectively.

\section{Quantization of constrained systems}

For the quantization of constrained systems we can use the
Dirac's method of quantization [1,2], or the path integral
quantization method [8-10].

Now will shall give a brief information about these two methods.
\subsection{Operator quantization}

For the Dirac's quantization method we have
\begin{equation}
H^{'}_{\a}\Psi=0,\;\;\;\a=0,n-r+1,...,n,
\end{equation}
where $\Psi$ is the wave function. The consistency conditions are
\begin{equation}
[H'_{\m}, H'_{\n}]\Psi=0,\;\;\;\m,\n=1,...,r,
\end{equation}
where$[,]$ is the commutator. The constraints $H'_{\a}$ are
called first- class constraints if they satisfy
\begin{equation}
[H'_{\m}, H'_{\n}]=C_{\m\n}^{\g}H'_{\g}.
\end{equation}

In the case when the Hamiltonians $H'_{\m}$ satisfy
\begin{equation}
[H'_{\m}, H'_{\n}]=C_{\m\n},
\end{equation}
with $C_{\m\n}$ do not depend on $q_{i}$ and $p_{i}$, then from
(8) there arise naturally Dirac' brackets and the canonical
quantization will be performed taking Dirac's brackets into
commutators.

\subsection{Path integral quantization method}

The path integral quantization is an alternative method to perform
the quantization of constrained systems.

Now we shall give a brief review of the canonical path integral
formulation of constrained systems [8-10].

If the set of equations (4) is integrable then one can solve them
to obtain the canonical phase-space coordinates as
\begin{equation}
q_{a}\equiv q_{a}(t, t_{\m}),\;\;\;p_{a}\equiv p_{a}(t,
t_{\m}),\;\;\m=1,...,r,
\end{equation}
In this case, the path integral representation may be written as
[8-10]

\bn &&\langle Out|S|In\rangle=\int \prod_{a=1}^{n-r}dq^{a}~dp^{a}
\exp [i \{\int_{t_{\a}}^{{t'}_{\a}}(-H_{\a}+ p_a\frac{\p
H^{'}_{\a}}{\p p_a})dt_{\a}\}],\nonumber\\&&a=1,...,n-r,
\;\;\;\a=0,n-r+1,...,n. \en

One should notice that the integral (12) is an integration over
the canonical phase - space coordinates $(q_a, p_a)$.

\section{Quantization of the relativistic particle in a plane wave}

Motion of spinless particle in the external field of a plane wave
is defined by a singular Lagrangians. The usual approach that
authors investigate is the Dirac's approach [1,2].

Let us consider the reparametrization invariant action of a
spinless particle in the external electromagnatic field of a
plane wave directed a long the axis $x^{3}$ as

\bn S=&& -\int (m\sqrt{{\dot{x}}^{2}} + e{\dot {x}}A)d\tau,\\
=&& -\int [m\sqrt{2{\dot {x}}_{+}{\dot{x}}_{-}-({\dot{x}}_{\bot})^{2}} + e{\dot {x}^{a}}A_{a}(x_{-})]d\tau,\nonumber\\
A_{\m}=&& (0, A^{a}, 0),\;\; x_{\pm} = (x^{0} {\pm}
x^{3})/\sqrt{2}, \;\;\; x_{\bot}=(x^{a}),\;a=1, 2.\nonumber \en

The momenta conjugated to $x_{\pm}$ and $x^{a}$ are \bn
&&\pi_{+}= \frac{\p L}{\p \dot{x}_{+}}= -\frac{m
{\dot{x}}_{-}}{\sqrt{2{\dot
{x}}_{+}{\dot{x}}_{-}-({\dot{x}}_{\bot})^{2}}},\\
&&\pi_{-}= \frac{\p L}{\p \dot{x}_{-}}= -\frac{m
{\dot{x}}_{+}}{\sqrt{2{\dot
{x}}_{+}{\dot{x}}_{-}-({\dot{x}}_{\bot})^{2}}},\\
&&\pi_{a}=  \frac{\p L}{\p \dot{x}^{a}}= \frac{m
{\dot{x}}^{a}}{\sqrt{2{\dot
{x}}_{+}{\dot{x}}_{-}-({\dot{x}}_{\bot})^{2}}} -e A_{a}(x_{-}),
\en the primary constraint is in the form
\begin{equation}
\f^{(1)}= \pi_{-} - \frac{[\pi_{a} +e A_{a}(x_{-})]^{2} +
m^{2}}{2\pi_{+}} =0,\;\;\pi_{\pm}\neq0.
\end{equation}
The canonical Hamiltonian (3) vanishes identically. So the primary
Hamiltonian is the constraint itself multiplied by a function
$\la(\tau)$, i.e
\begin{equation}
H^{(1)}= \la {\f}^{(1)}.
\end{equation}
The equations of motion read as \bn&&{\dot{x}}_{+}= \{x_{+},
H^{(1)}\}= \la \frac{[\pi_{a} +e A_{a}(x_{-})]^{2} +
m^{2}}{\pi_{+}},\\
&&{\dot{x}}_{-}= \{x_{-}, H^{(1)}\}= \la,\\
&&{\dot{x}}_{a}= \{x_{a}, H^{(1)}\}= \la \frac{(\pi_{a} +e
A_{a}(x_{-}))}{\pi_{+}},\\
&&{\dot {\pi}}_{+}=\{\pi_{+},H^{(1)}\}=0,\\
&&{\dot {\pi}}_{a}=\{\pi_{a},H^{(1)}\}=0,\\
&&{\dot {\pi}}_{-}=\{\pi_{-},H^{(1)}\}=-\la e \frac{(\pi_{a} +e
A_{a}(x_{-}))}{\pi_{+}}\frac{\p A_{a}}{\p x_{-}},\en

Since we have only one first class constraint, the theory is
degenerate and it is impossible to find $\la$ in the Dirac's
procedure. To determine $\la$, one should impose time-dependent
gauge fixing of the  form
\begin{equation}
\Phi^{G}= x_{-}- \tau=0.
\end{equation}
From the condition of conservation of this gauge in the time
$\tau$
\begin{equation}
\frac{d\Phi^{G}}{d\tau}=\frac{\p \Phi^{G}}{\p \tau} + \{\Phi^{G},
H^{(1)}\}=(\la- 1)=0,
\end{equation}
we obtain $\la=1$. No other constraints arise.

Now we would like to study the quantization of the same problem
using the canonical method [4-7]. Equations (14) and (16) lead us
to obtain the expressible velocities $ {\dot {x}}_{+}$ and $ {\dot
{x}}_{a}$ in terms of the primary unexpressible velocity $ {\dot
{x}}_{-}$ as

\begin{equation}
{\dot {x}}_{+} =\frac{\pi_{-}}{\pi_{+}}{\dot {x}}_{-}=w_{+}
,\;\;{\dot {x}}_{a} =\frac{(\pi_{a}+ eA_{a}) }{\pi_{+}}{\dot
{x}}_{-}=w_{a}.
\end{equation}
Substituting (27) in (15), one gets

\begin{equation}
\pi_{-} = \frac{[\pi_{a} +e A_{a}(x_{-})]^{2} + m^{2}}{2\pi_{+}}
=- H_{x_{-}},\;\;\pi_{\pm}\neq0.
\end{equation}

The canonical Hamiltonian $H_{\tau}$ can be written as
\begin{equation}
H_{\tau}= -L + \pi_{a}w_{a} + \pi_{+}w_{+} - {\dot
{x}}_{-}H_{x_{-}}.
\end{equation}
Explicit calculations show that $H_{\tau}$ vanishes identically.

Making use of equations (2) and (28,29), one obtains the set of
Hamilton Jacob partial differential equations as
\bn&&{H'}_{\tau}=\pi_{\tau}=0,\;\;\;\;\;\;\;\;\;\;\;\;\;\;\;\;\;\;\;\pi_{\tau}=\frac{\p S}{\p \tau},\\
&&{H'}_{x_{-}} = \pi_{-} + H_{x_{-}}=0,\;\;\;\;\pi_{-} = \frac{\p
S}{\p x_{-}}.\en The equations of motion are obtained as total
differential equations in many variables as follows: \bn
dx_{+}&&=\frac{\p {H'}_{\tau}}{\p \pi_{+}}d\tau + \frac{\p
{H'}_{x_{-}}}{\p \pi_{+}}dx_{-},\nonumber\\
&&=\frac{[\pi_{a} +e A_{a}(x_{-})]^{2} + m^{2}}{\pi_{+}}dx_{-},\\
dx_{a}&&=\frac{\p {H'}_{\tau}}{\p \pi_{a}}d\tau + \frac{\p
{H'}_{x_{-}}}{\p \pi_{a}}dx_{-},\nonumber\\
&&=\frac{(\pi_{a} +e
A_{a}(x_{-}))}{\pi_{+}}dx_{-},\\
d\pi_{+}&&=-\frac{\p {H'}_{\tau}}{\p x_{+}}d\tau - \frac{\p
{H'}_{x_{-}}}{\p x_{+}}dx_{-}=0,\\
d\pi_{a}&&=-\frac{\p {H'}_{\tau}}{\p x_{a}}d\tau - \frac{\p
{H'}_{x_{-}}}{\p x_{a}}dx_{-}=0,\\
d\pi_{-}&&=-\frac{\p {H'}_{\tau}}{\p x_{-}}d\tau - \frac{\p
{H'}_{x_{-}}}{\p x_{-}}dx_{-}=0,\\
&&=- e \frac{(\pi_{a} +e A_{a}(x_{-}))}{\pi_{+}}\frac{\p
A_{a}}{\p x_{-}}dx_{-},\\
d\pi_{\tau}&&=-\frac{\p {H'}_{\tau}}{\p \tau}d\tau - \frac{\p
{H'}_{x_{-}}}{\p \tau}dx_{-}=0.\en

In order to have a consistent theory, one should consider the
variations of constraints in general. Since the total variations
of the constraints ${H'}_{\tau}$ and  ${H'}_{x_{-}}$ are
identically zero, no further constraints arise and the equations
of motion are integrable. Hence, the canonical phase space
coordinates $(x_{+}, \pi_{+})$ and $(x_{a}, \pi_{a})$ are
obtained in terms of parameters $\tau$ and $x_{-}$.

To obtain the operator quantization of this system one can follow
the procedure discussed in section (2.1). In this case one takes
the constraint equation as an operator whose action on the allowed
Hilbert space vectors is constrained to $zero$, i., e., $
{H'}_{x_{-}}\Psi =0$, we obtain
\begin{equation}
i\frac{\p \Psi}{\p x_{-}}=- \frac{[-i\frac{\p}{\p x^{a}} -e
A^{a}(x_{-})]^{2} + m^{2}}{2\pi_{+}}\Psi
\end{equation}

One should notice that (39) is the Klein-Gordon equation. In fact,
the Klein-Gordon equation
\begin{equation}
({\mathcal{P}}^{2} - m^{2})\Psi(x)=0,\;\;{\mathcal{P}}_{\m}=
i\p_{\m} - e A_{\m}(x),
\end{equation}
in the external field of a plane wave (13), being written in the
light cone variables $(x_{\pm}, x_{\bot}=(x^{a}))$, reads as
\begin{equation}
(2\frac{\p^{2}}{\p x_{-} \p x_{+}} + [ -i \frac{\p}{\p x^{a}}- e
A^{a}(x_{-})]^{2} + m^{2})\Psi(x)=0.
\end{equation}

 Now to obtain the path integral quantization of this system,
we can use equation (5) to obtain the canonical action as
\begin{equation}
S=\int(\frac{[\pi_{a} +e A_{a}(x_{-})]^{2} + m^{2}}{2\pi_{+}}
+\pi_{+}{\dot {x}}_{+} + \pi_{a}{\dot {x}}_{a})dx_{-}.
\end{equation}
Making use of (42) and (12) the path integral for the system (13)
is obtained as \bn &&\langle x_{+}, x_{a}, \tau, x_{-}; {x'}_{+},
{x'}_{a}, {\tau}^{'}, {x'}_{-}\rangle= \int_{x_{+},
x^{a}}^{{x'}_{+}, {x'}_{a}}\prod
dx_{+}~dx_{a}~d\pi_{+}~d\pi_{a}\nonumber\\&&\exp
[i\{\int_{x_{-}}^{x_{-}'}(\frac{[\pi_{a} +e A_{a}(x_{-})]^{2} +
m^{2}}{2\pi_{+}} +\pi_{+}{\dot {x}}_{+} + \pi_{a}{\dot
{x}}_{a})dx_{-}\}]. \en

In fact, the path integral representation (43) is an integration
over the canonical phase-space coordinates  This path integral
representation is an integration over the canonical phase space
coordinates $(x_{+}, \pi_{+}, x_{a}, \pi_{a})$.

Now for a system with $ n$ degrees of freedom  and $ r$ first
class constraints $\f^{\a}$, the matrix element of the $S$ -
matrix is given by Faddeev and Popov [12,13] as
\begin{equation}
\langle Out|S|In\rangle=\int \prod_{t} d{\m}(q_j, p_j)\exp [i
\{\int_{-\infty}^{\infty}dt(p_j \dot{q_j} - H_0)\}],\;\;j=1,...,n,
\end{equation}
where the measure of integration is given as
\begin{equation}
d {\m}= det|\{\f^{\a},
\k^{\bt}\}|\prod_{\a=1}^{r}\delta(\k^{\a})\delta(\f^{\a})\prod_{j=1}^{n}dq^{j}
dp_{j},
\end{equation}
and $\k_{a}$ are $r$- gauge constraints.

If we perform now the usual path integral quantization [12,13]
using (44) for the system (13), one must choose one time dependent
gauge fixing (equation (25)) condition to obtain the path integral
quantization over the canonical phase-space coordinates.

\section{ Conclusion}

Path integral quantization for relativistic particle in plane
wave is obtained by using the canonical path integral formulation
[8-10]. In this approach, since the integrability conditions
$d{H'}_{\tau} =0$ and  $d{H'}_{x_{-}}=0$ are satisfied
identically, this system is integrable. Hence, the canonical
phase-space coordinates $(x_{+}, \pi_{+}, x_{a}, \pi_{a})$ are
obtained in terms of parameter $(x_{-})$. In this case the path
integral, then follows directly as given in (43) without using any
gauge fixing conditions.

When applying the analog of Faddeev's method [12,13] to this
model, one may choose gauge fixing of the form $x_{-} -\tau =0$,
so that one can integrate over the extended phase space
coordinates $(x_{-}, \pi_{-}, x_{+}, \pi_{+}, x_{a}, \pi_{a})$
and after integration over the redundant variables $(x_{-},
\pi_{-})$, one can arrive at the result (43). Besides different
choice of gauge fixing will lead to different quantum theories.
In this case there are many measures one can use to define an
integral. In other words considerable care must be taken with the
choice of gauge fixing conditions in order to arrive at the
correct result.

As a conclusion, the path integral for relativistic particle in a
plane wave is obtained without using any gauge fixing conditions
and without the problems of the measure. Besides the canonical
action integral is obtained with the consequncent constraints
(30,31) are enforced explicitly without Lagrange multipliers.

\end{document}